\documentclass[3p,times,procedia]{elsarticle}
\usepackage{nupha_ecrc}
\usepackage{lineno}

\volume{00}

\firstpage{1}

\journalname{Nuclear Physics A}

\runauth{}

\jid{nupha}

\jnltitlelogo{Nuclear Physics A}

\usepackage{amssymb}

\usepackage[figuresright]{rotating}

\begin{document}
\begin{frontmatter}




\title{Investigating the Role of Coherence Effects on Jet Quenching in Pb-Pb Collisions at $\sqrt{s_{NN}} =2.76$ TeV using Jet Substructure }


\author{Nima Zardoshti \\
On behalf of the ALICE Collaboration \\
(University of Birmingham)}

\address{}

\begin{abstract}
We report measurements of two jet shapes, the ratio of 2-Subjettiness to 1-Subjettiness ($\it{\tau_{2}}/\it{\tau_{1}}$) and the opening angle between the two axes of the 2-Subjettiness jet shape, which is obtained by reclustering the jet with the exclusive-$\it{k}_{T}$ algorithm~\cite{kT} and undoing the final clustering step. The aim of this measurement is to explore a possible change in the rate of 2-pronged objects in Pb-Pb compared to pp due to colour coherence. Coherence effects~\cite{ColourCoherence} relate to the ability of the medium to resolve a jet's substructure, which has an impact on the energy loss magnitude and mechanism of the traversing jet. In both collision systems charged jets are found with the anti-$\it{k}_{T}$ algorithm~\cite{Anti-kT}, a resolution parameter of $\it{R}=$0.4\ and a constituent cut off of $0.15$ GeV. This analysis uses hadron-jet coincidence techniques in Pb-Pb collisions to reject the combinatorial background and corrects further for background effects by employing various jet shape subtraction techniques and two dimensional unfolding. Measurements of the Nsubjettiness for jet momenta of $40-60$ GeV/$\it{c}$ in Pb-Pb collisions at $\sqrt{s_{NN}} =2.76$ TeV and pp collisions at $\sqrt{s} =7$ TeV will be presented and compared to PYTHIA simulations.
\end{abstract}




\end{frontmatter}


\section{Introduction}
Jets are collimated sprays of hadrons resulting from the fragmentation of partons scattered with high virtuality $\it{Q^{2}}$ in the initial stages of the collision. Identifying and clustering these hadrons and reconstructing the jet, provides the best experimentally accessible proxy for the initial scattered parton. In relativistic heavy ion collisions the partons are scattered before the formation of the Quark-Gluon-Plasma (QGP), making them a natural probe to study the medium. Partons traversing the QGP interact and lose energy to the medium, via radiative and collisional processes, which is expected to modify the final state hadron distribution. Modifications in the distribution of jet constituents, in Pb-Pb collisions relative to pp, can provide information on the parton's interactions with the medium. Variables known as jet shapes, which characterise this distribution of hadrons in a jet, are an experimental tool used to pursue this goal.

One of the recently discussed ingredients of the parton-medium interaction is colour coherence, determined by a transverse resolution scale. Partons traversing the QGP separated by distances larger than this scale are thought to be resolved as independent colour charges and interact incoherently with the medium, whereas partons separated by smaller angles interact coherently with the QGP and radiate as a colour singlet. This coherence resolution scale can be characterised with a critical angle, $\it{\theta_{c}}$, which is related to the medium's characteristic scale, $\Delta_{Med}$ via Eq~\ref{Eq_ColourCorrelationDistance}. Here $\hat q$ is the medium transport parameter, L is the medium's length and $r_{\perp}$ is the jet's transverse extension in the medium. $\it{\theta}$ is the jet's opening angle, which is given by the largest antenna (the two substructures with the largest angular separation) in the jet which in a vacuum corresponds to the first splitting. Recent theoretical work has highlighted the sensitivity of two-pronged jets to coherence effects in the QGP~\cite{TwoProngProb}. Therefore, constructing jet shapes measuring the two-prongness of jets, using new unclustering techniques to find the relevant jet antenna, can be an experimental method of measuring coherence effects.

\begin{equation}
 \Delta_{med} \simeq 1-e^{-\frac{1}{12}\hat q Lr^{2}_{\perp}} \equiv 1 - e^{-(\it{\theta}/\it{\theta_{c}})^{2}}.\\
\label{Eq_ColourCorrelationDistance}
\end{equation}

\section{Jet Shapes}

Nsubjettiness, $\it{\tau_{N}}$, is a jet shape initially developed to tag two-pronged Higgs decays~\cite{NSubjettiness}. $\it{\tau_{N}}$, where $\it{N}$ can be any positive integer, is a measure of how $\it{N}$-cored a jet's substructure is and is defined by Eq~\ref{Eq_Nsubjetiness}, where the subscript $\it{i}$ denotes each track in the jet and $\it{N}$ is the number of axes found in the jet. By construction, a jet with a $\it{\tau_{N}}$ value approaching $0$ is said to have $\it{N}$ or fewer definite cores, whereas a jet that has a $\it{\tau_{N}}$ value approaching unity has at least $\it{N}+1$ substructures. Therefore the ratio of $\it{\tau_{N}}/\it{\tau_{N-1}}$ is sensitive to exactly $\it{N}$ cores in the jet. It follows that $\it{\tau_{2}}/\it{\tau_{1}}$ is a jet shape sensitive to $2$-pronged jets. The aperture angle between the two axes in the calculation of $\it{\tau_{2}}$ is also measured and denoted by $\Delta \it{R}$. This is represented by Eq~\ref{Eq_DeltaR}. These axes are obtained by reclustering the jet using the exclusive-$\it{k}_{T}$ algorithm and unwinding the last step in the reclustering procedure. 

\begin{equation}
 \it{\tau_{N}}=\frac{\sum_{i=1}^{\it{N}}p_{T,i} min(\Delta \it{R}_{1,i},\Delta \it{R}_{2,i},...,\Delta \it{R}_{N,i})}{\sum_{i=1}^{N}(p_{T,i}\it{R})} \\
\label{Eq_Nsubjetiness}
\end{equation}

\begin{equation}
\Delta \it{R}_{a,b}=\sqrt{(\it{\eta}_{axis,a}-\it{\eta}_{axis,b})^{2}+(\it{\varphi}_{axis,a}-\it{\varphi}_{axis,b})^{2}}. \\
\label{Eq_DeltaR}
\end{equation}

\section{Analysis Procedure}

\begin{figure}[h!]
	\centering
		\includegraphics[width=0.55\textwidth]
{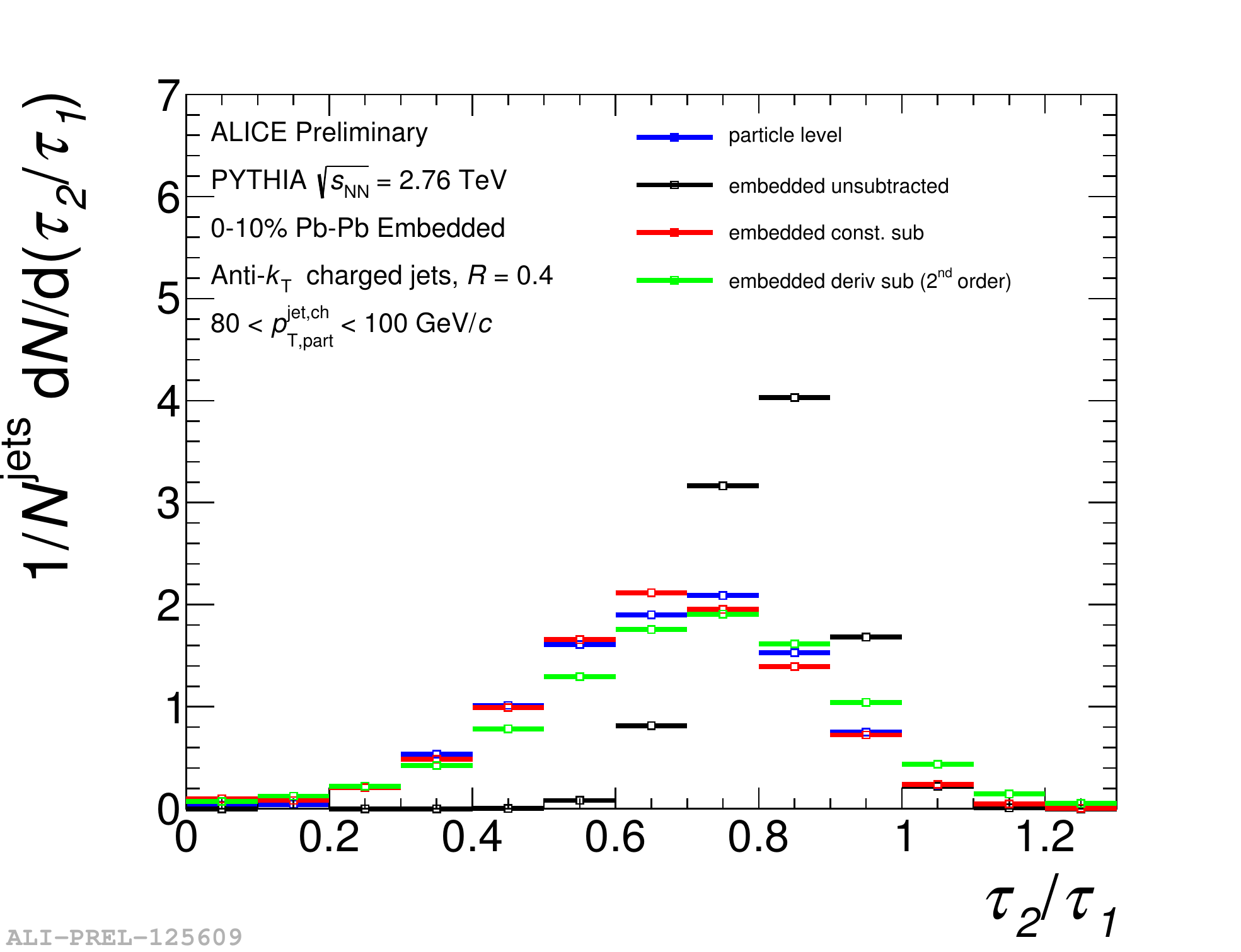}
\caption{Performance of background subtraction techniques in Pb-Pb collisions for the $\it{\tau_{2}}/\it{\tau_{1}}$ jet shape}
\label{Tau2to1_Background}
\end{figure}

Two collision systems were studied in this analysis. In pp, a minimum bias sample of events at $\sqrt{s}=7$ TeV was selected and in Pb-Pb the $0-10\%$ most central events were chosen at $\sqrt{s_{NN}}=2.76$ TeV. Jet finding was performed on charged particles with a constituent cut-off $\it{p}_{T}^{ch}$=150 MeV/$\it{c}$, using the anti$-\it{k}_{T}$ algorithm with resolution $\it{R}=$0.4 using the FastJet package~\cite{FastJet}. The constituents in the jet were recombined using the E-scheme. The antenna axes used for both jet shapes were found using the exclusive-$\it{k}_{T}$ algorithm. In Pb-Pb the background subtraction techniques of constituent subtraction~\cite{ConstSub} (default) and second order derivative subtraction~\cite{DerivSub} were used to remove the average uncorrelated background from the shape observables. The performance of the subtraction methods are shown in Fig~\ref{Tau2to1_Background}. Particle level PYTHIA jets were generated and the detector response was then simulated with a full detector simulation in GEANT3. These detector level jets were then embedded into real $0-10 \%$ most central Pb-Pb events. These jets were then reconstructed in the presence of the heavy-ion background (embedded level) and matched to the detector level PYTHIA jets which were in turn matched to the particle level PYTHIA jets. This suppresses the contamination of fake jets in the sample. The heavy-ion background shifts the jet shape to larger values (black), compared to the original PYTHIA jet (blue). However both background subtraction techniques (red and green) perform well at removing the average uncorrelated background and correcting the variable back to its true value. 

In Pb-Pb data, a 2D extension to the hadron-jet coincidence technique~\cite{RecoilJets} was also used to reject combinatorial jets from the measured sample. This method measures the difference in the yield of jets recoiling from two high $\it{p}_{T}^{ch}$ trigger hadron classes. The trigger hadron $\it{p}_{T}^{ch}$ classes used were $15-45$ GeV/$\it{c}$ (signal class) and $8-9$ GeV/$\it{c}$ (reference class). For each class, jet finding was performed in an angular window of $| \pi - \Delta\varphi | <$ 0.6, where $ \Delta\varphi$ was the difference in $\varphi$ angle between the trigger hadron and the jet axis. In both collision systems a 2D Bayesian unfolding procedure was used to correct both the $\it{p}_{T}^{jet,ch}$ and shape observable simultaneously for background fluctuations and detector effects. Fully corrected results are reported in the range $40 < \it{p}_{T}^{jet,ch} <$ 60 GeV/$\it{c}$ for both systems.

\section{Results}

\begin{figure}[h!]
\centering
\begin{minipage}{.49\textwidth}
  \centering
  \includegraphics[width=1.07\linewidth]{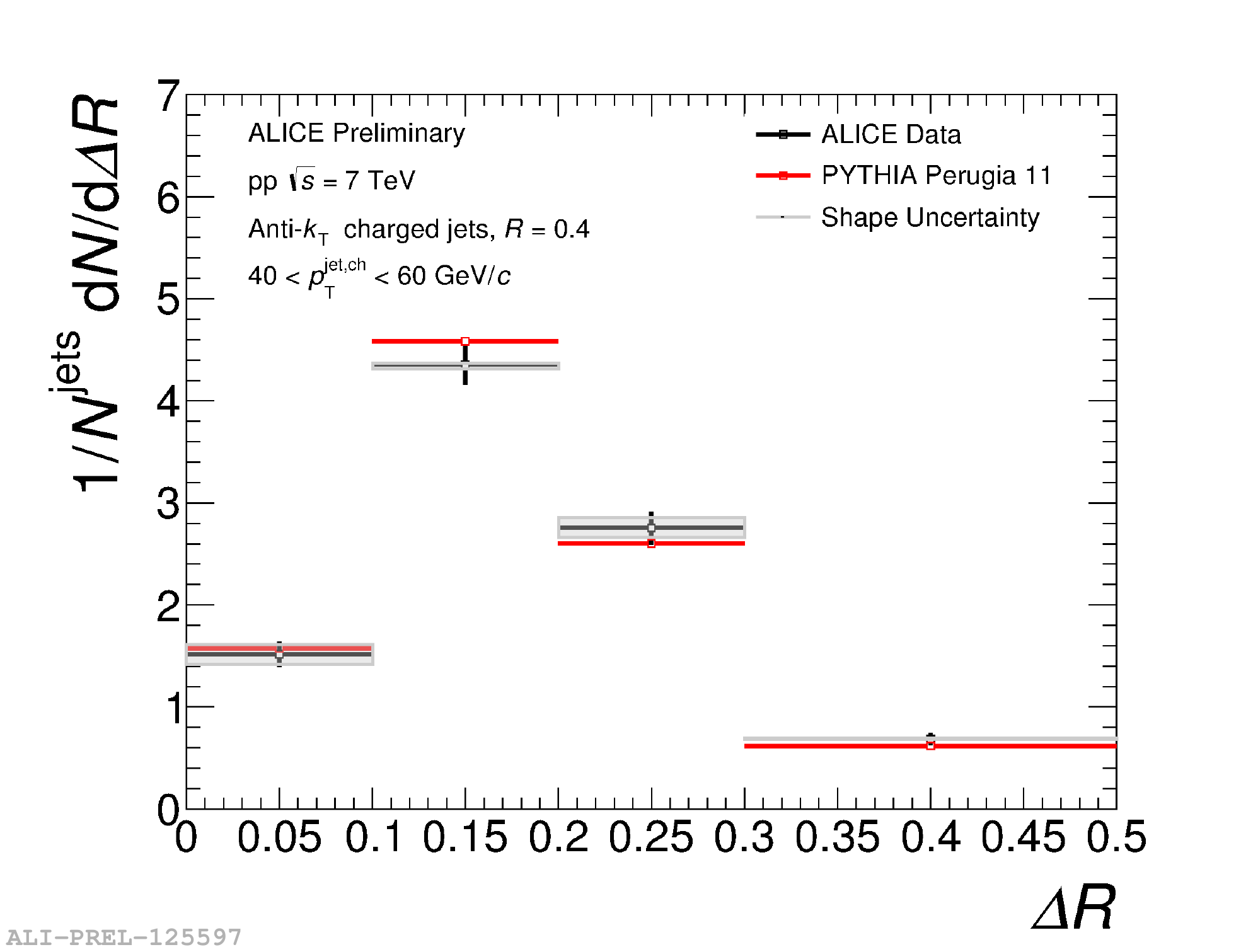}
\end{minipage}
\begin{minipage}{.49\textwidth}
  \centering
  \includegraphics[width=0.9\linewidth]{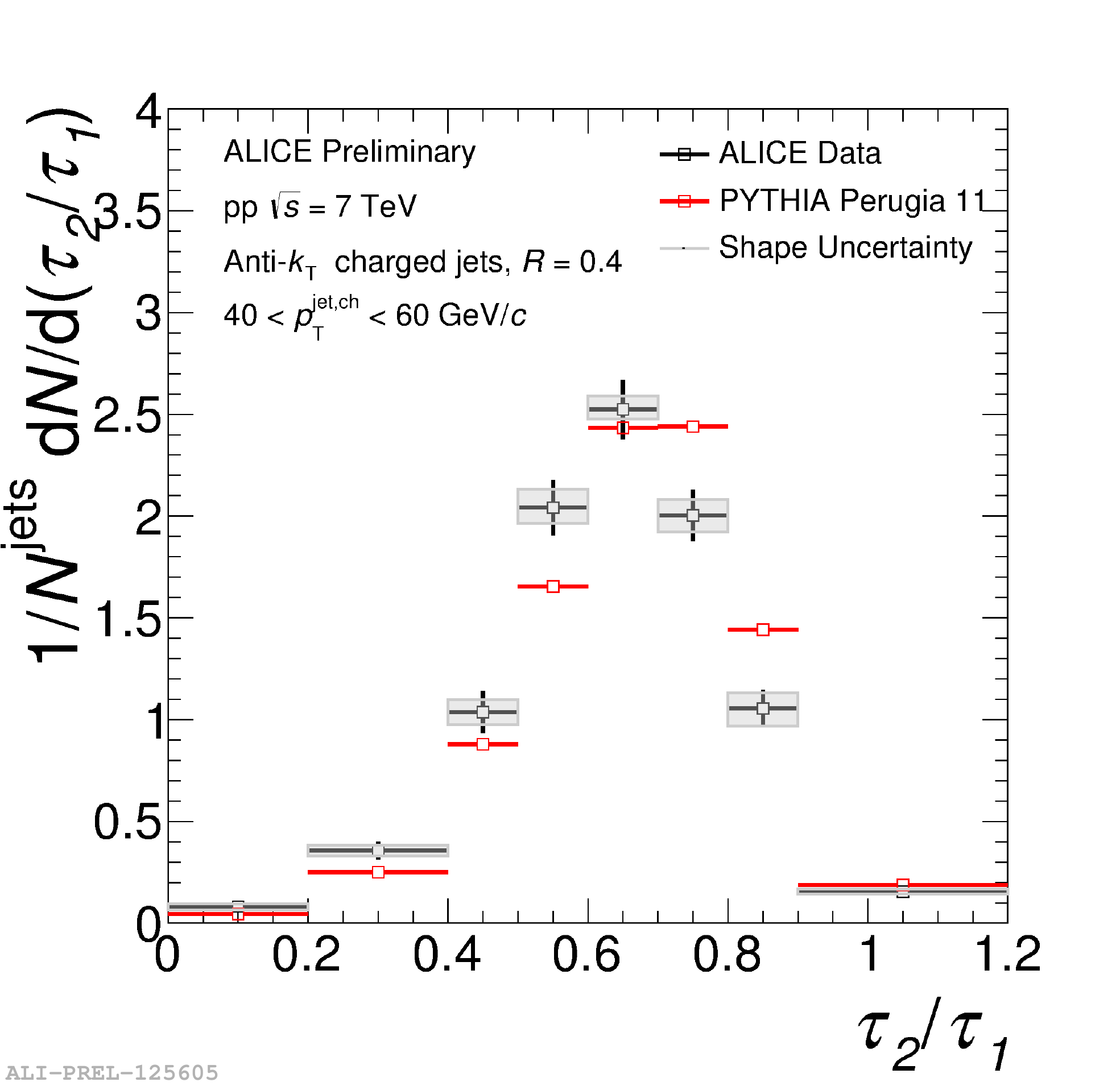}
\end{minipage}%
\caption{Fully corrected jet shapes in pp}
\label{JetShapes_pp}
\end{figure}

\begin{figure}[h!]
\centering
\begin{minipage}{.5\textwidth}
  \centering
  \includegraphics[width=.9\linewidth]{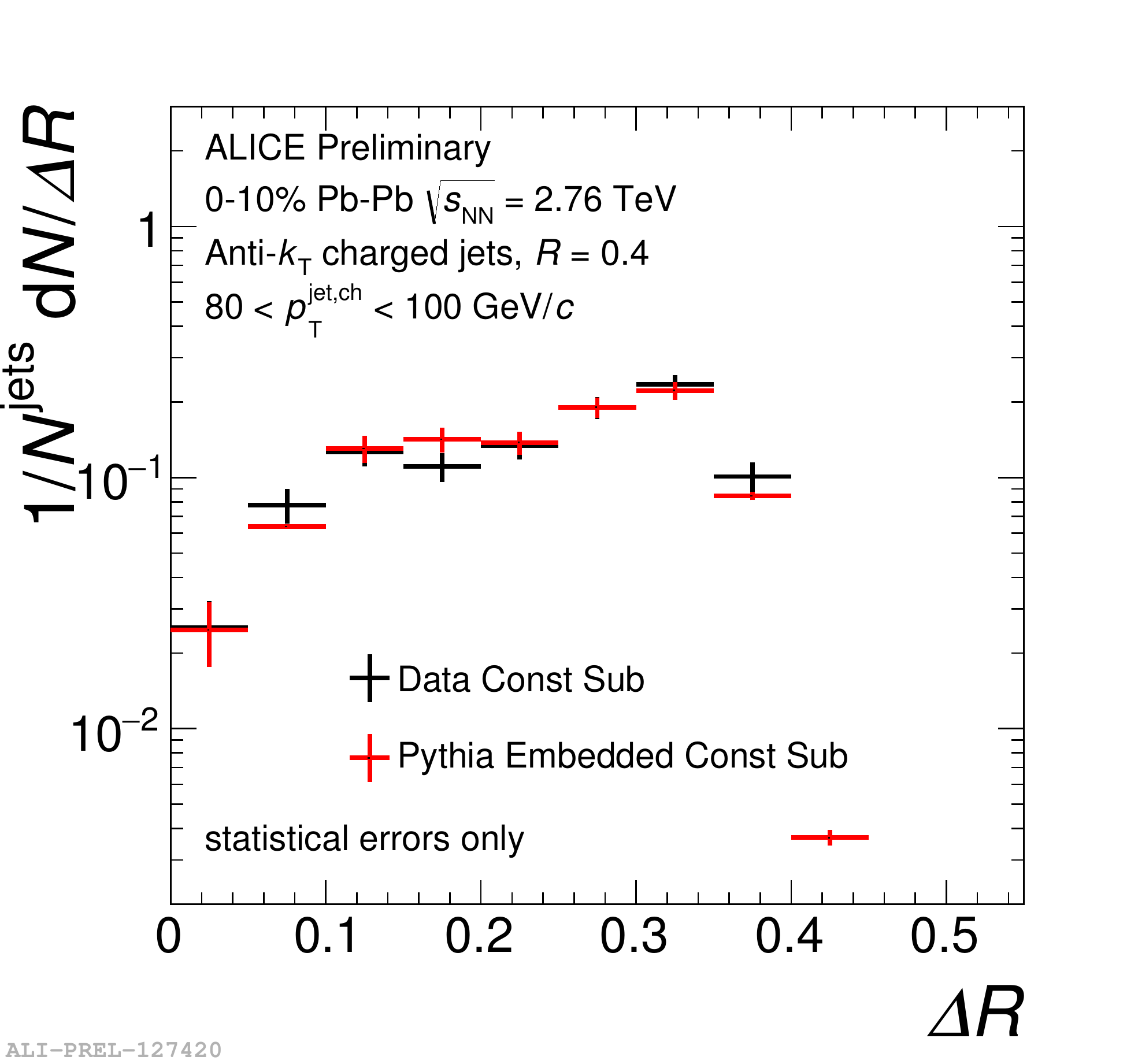}
\end{minipage}%
\begin{minipage}{.5\textwidth}
  \centering
  \includegraphics[width=1.07\linewidth]{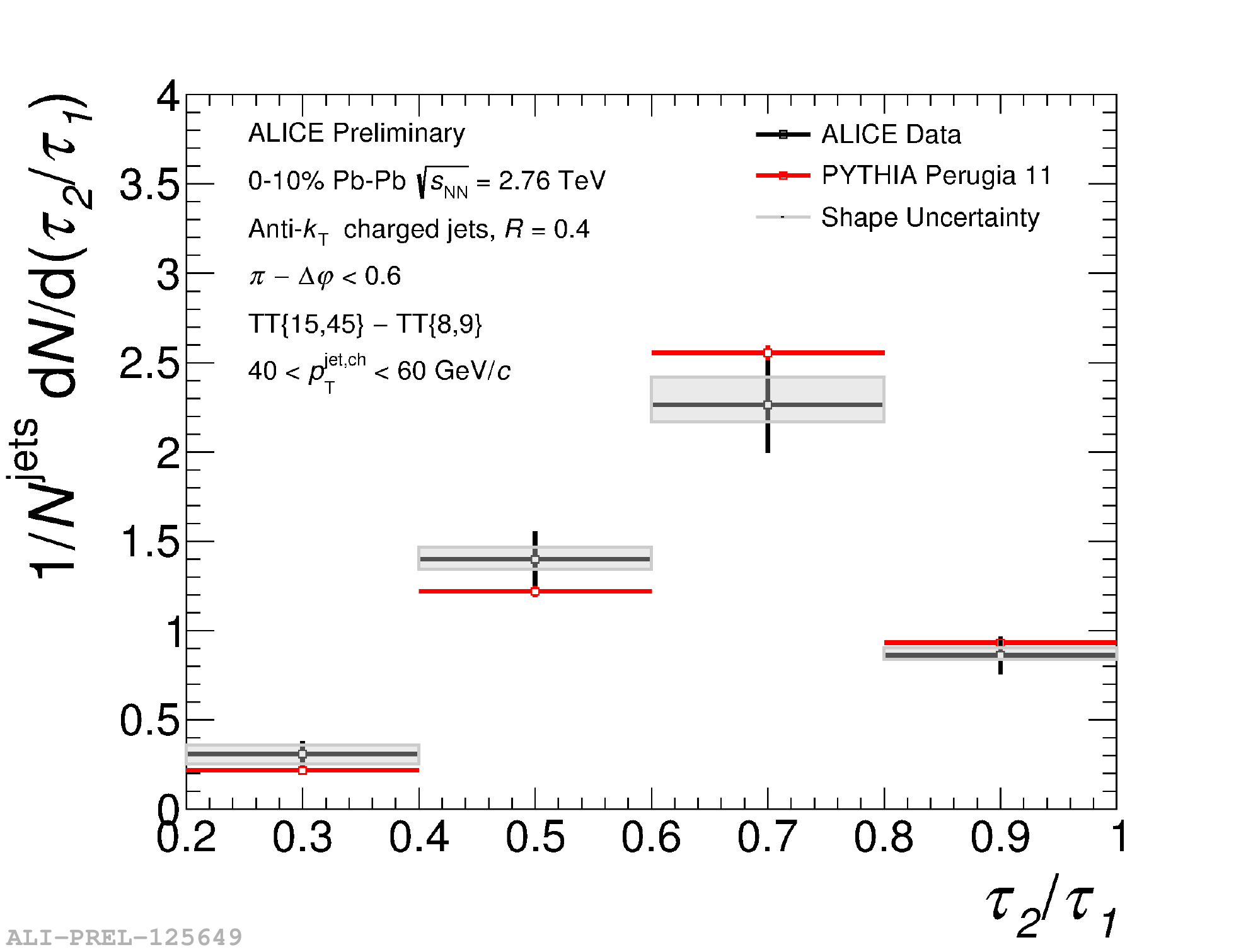}
\end{minipage}
\caption{Comparison of Pb-Pb inclusive vs PYTHIA embedded jets for $\Delta \it{R}$ (left) and Fully corrected results for $\it{\tau_{2}}/\it{\tau_{1}}$ in Pb-Pb (right)}
\label{JetShapes_PbPb}
\end{figure}

Fully corrected results in pp in the range $40 < \it{p}_{T}^{jet,ch} <$ 60 GeV/$\it{c}$ for the jet shapes $\it{\tau_{2}}/\it{\tau_{1}}$ and $\Delta \it{R}$ are shown in Fig~\ref{JetShapes_pp}. Here a comparison to PYTHIA~\cite{PYTHIA} Perugia 11 results at the same energy is also provided. It can be seen that PYTHIA describes the two axes ($\Delta \it{R}$) well, but there is slightly poorer agreement regarding the fragmentation around those two axes ($\it{\tau_{2}}/\it{\tau_{1}}$). In Pb-Pb data, full corrections for the $\Delta \it{R}$ variable using the hadron-jet coincidence technique (recoil jets) are still ongoing, however a comparison of the inclusive measured data to the embedded level is provided in Fig~\ref{JetShapes_PbPb} at a higher $\it{p}_{T}^{jet,ch}$ where combinatorial jets are suppressed. The agreement between the two distributions leaves little room for quenching effects in the data. However, it needs to be noted that large values of $\Delta \it{R}$ are dominated by a fake second axis due to unsubtracted remnants of the heavy ion background. Therefore, large corrections to the variable are expected via the unfolding procedure. The variable $\it{\tau_{2}}/\it{\tau_{1}}$ shows less sensitivity to this effect and corrections are small (Fig~\ref{Tau2to1_Background}).

Figure~\ref{JetShapes_PbPb} shows the fully corrected $\it{\tau_{2}}/\it{\tau_{1}}$ jet shape in Pb-Pb in the range $40 < \it{p}_{T}^{jet,ch} <$ 60 GeV/$\it{c}$ using the hadron-jet coincidence technique. A comparison to PYTHIA Perugia 11 at the same energy using the hadron-jet coincidence technique is also provided. It can be seen that PYTHIA describes the data well, leaving little room for quenching modifications and coherence effects.


\section{Conclusions}

Fully corrected results for the two jet shapes, $\it{\tau_{2}}/\it{\tau_{1}}$ and $\Delta \it{R}$, are presented in the $40 < \it{p}_{T}^{jet,ch} <$ 60 GeV/$\it{c}$ range in pp collisions. Comparisons to PYTHIA show that the shapes are fairly well described by the model. In Pb-Pb data, the first fully corrected measurement of a jet shape ($\it{\tau_{2}}/\it{\tau_{1}}$) at large jet resolution parameter ($\it{R}=$ 0.4) and low $\it{p}_{T}^{jet,ch}$ ($40 < \it{p}_{T}^{jet,ch} <$ 60 GeV/$\it{c}$) is shown using the hadron-jet coincidence technique. Comparison to PYTHIA shows that the model also describes the Pb-Pb recoil data fairly well and no significant modifications due to coherence are discernible from this variable. Systematic investigation of the effects of different axis finding techniques is ongoing.





\bibliographystyle{elsarticle-num}
\bibliography{<your-bib-database>}



\end{document}